\documentclass{article}

\usepackage{arxiv}

\usepackage[utf8]{inputenc} % allow utf-8 input
\usepackage[T1]{fontenc}    % use 8-bit T1 fonts
\usepackage{hyperref}       % hyperlinks
\usepackage{url}            % simple URL typesetting
\usepackage{booktabs}       % professional-quality tables
\usepackage{amsfonts}       % blackboard math symbols
\usepackage{nicefrac}       % compact symbols for 1/2, etc.
\usepackage{microtype}      % microtypography
\usepackage{cleveref}       % smart cross-referencing
\usepackage{graphicx}
\usepackage{placeins}
\usepackage{natbib}
\usepackage{doi}
\usepackage[subrefformat=parens]{subcaption}
\usepackage{ragged2e}

\title{microAI: A machine learning tool for fast calculation of lift coefficients in microchannels}

 \date{}

\author{ 
	Erfan Hamdi\thanks{Same Contribution}\\
	Department of Mechanical Engineering\\
	Sharif University of Technology\\
	Tehran, Azadi St. \\
	\texttt{erfan.hamdi@mech.sharif.ir} \\
	\And
	Rasool Dezhkam$^{*}$ \\
	Department of Mechanical Engineering \\
	Sharif University of Technology\\
	Tehran, Azadi St.\\
	\texttt{rasool.dezhkam@sharif.edu} \\
	\AND
	Amir Shamloo\thanks{Corresponding Author}\\
	Department of Mechanical Engineering\\
	Sharif University of Technology\\
	Tehran, Azadi St.\\
	\texttt{shamloo@sharif.ir} \\
	\And
	Ali Mashhadian\\
	Department of Mechanical Engineering\\
	Sharif University of Technology\\
	Tehran, Azadi St.\\
	\texttt{alimashhadian77@gmail.com} \\
}

\renewcommand{\headeright}{}
\renewcommand{\undertitle}{}
\renewcommand{\shorttitle}{microAI: A machine learning tool for fast calculation of lift coefficients in microchannels}

\begin{document}
\maketitle

\begin{abstract}
	There have been multiple methods proposed to calculate lift coefficients in microfluidic channels. One of the most used methods is using Direct Numerical Simulation. DNS is a very accurate yet computationally expensive method. DNS computations comprise most of the time consumed on a microfluidic simulation done by commercial software. This paper proposes a user-friendly, fast, and accurate AI-based webapp named microAI that can calculate the microfluidic lift coefficients of channels. We have also studied the effects of different types of activation functions and optimizers in convergence and the final function's differentiability. microAI is deployed to huggingface and is accessible at  \url{https://erfanhamdi.github.io/microAI/}
\end{abstract}

% keywords can be removed

\section{Introduction}
Inertial microfluidics as a method for cell separation is frequently used in Lab-On-a-Chip (LOC) and Lab-On-a-Disk (LOD) platforms. Inertial cell separation belongs to the passive category of cell separation methods. There is no external force for cell manipulation inside the microchannels, and particles will be separated because of differences in density, shape, etc. On the other hand, external forces are applied to the fluid or the particle in active methods such as dielectrophoresis \citep{kwizera2021methods}, magnetophoresis \citep{nasiri2022design}, acoustophoresis \citep{collins2017continuous}, etc. However, passive methods like deterministic lateral displacement (DLD) \citep{mcgrath2014deterministic}, Pinch Flow Fractionation (PFF) \citep{yamada2004pinched}, and inertial methods \citep{di2009inertial} unlike active methods,  do not apply any external force. Resulting in less complicated methods, they have always been one of the researchers' choices for particle sorting or separation.
Lift and drag forces are the two main forces in the passive methods. The drag force and the direction of the fluid streamline are the same, and the movement of particles is mainly affected by this force. However, the lift force is orthogonal to the flow, resulting from pressure and surface stress differences around the particle. This difference in lift force between various types of particles moves them to different positions in the width of the channel. 

Calculating the lift force applied to spherical particles has always been discussed \citep[]{saffman1965lift, asmolov1999inertial, ho1974inertial, bazaz2020computational}.
Consequently, the lift force consists of four terms: Saffman, Magnus, Wall-induced, and Shear gradient lift force \citep{zhang2016fundamentals}. The lag between the particles and the fluid made by wall effects \citep{saffman1965lift} induces Saffman force. Magnus force is the result of particle rotation. Pressure is higher on one side of a rotating particle than on the other side; therefore,  a lateral force is applied to the particle. Wall-induced lift force is because of the wall effect. The flow field changes near the wall because of the particle, and a velocity gradient is made, and as a result, a shear rate makes the particle rotate and move to the center of the channel to be far from the wall \citep{michaelides2006particles}. Shear gradient lift force is applied to the particles because of the parabolic form of the velocity profile inside the channels. Therefore, the relative velocity is different on the two sides of the particles. Thus, a shear gradient moves them toward the wall until wall-induced, and the shear gradient lift forces cancel each other \citep{feng1994direct}.

All of the mentioned lift forces have a closed form, which enables us to calculate them separately. However, Direct Numerical Solution (DNS) is a method that calculates the total lift force, comprised of all forms of the lift force depending on the particle size, particle position, and channel Reynolds number. This method, developed by Di Carlo et al., calculates the lift force in any arbitrary coordinate across the cross-section \citep{di2009particle}. As we used the DNS method in our previous work \citep{mashhadian2019inertial}, we concluded that although DNS is the most accurate method for inertial lift calculation, it is time-consuming and computationally expensive. Hence, utilizing Artificial Intelligence (AI) methods can help researchers have almost the same accuracy and reduce the computational cost \citep{su2021machine}.

Deep Learning is a class of Machine Learning methods that can learn Representations of the input data by stacking multiple layers of Perceptron \citep{lecun2015deep}. These methods have resulted in breakthroughs in many fields and tasks, such as image classification \citep{krizhevsky2017imagenet}, natural language processing \citep{brown2020language} and highly accurate prediction of protein structure \citep{jumper2021highly} which is a significant breakthrough in bioengineering.
Due to the high computational cost of conventional methods available in commercial packages, the iterative process of coming up with solutions for new problems has been experiencing difficulties in scientific fields. With faster processing units such as GPUs and more data, using AI as a surrogate model has become more feasible \citep{mcbride2019overview, mohammadzadeh2022predicting}. The progress made in high-throughput microfluidics has resulted in vast amounts of data. Processing this data using conventional methods no longer results in a good performance.

AI methods have been deployed in many microfluidic applications to address that need. In a great review, \citep{riordon2019deep}, have classified the ways that Deep Learning has been used in microfluidic applications based on the type of input data and the desired output data from unstructured-to-unstructured \citep{chen2016deep} which cell classification is a type of, to Image-to-Image types of application for cell segmentation \citep{zaimi2018axondeepseg}. Design and controlling of the microfluidic devices is an expert dependent process which can be of a burden to wide adoption of microfluidics in other scientific fields. This gap can also be bridged by usage of user-friendly and easy to use machine learning methods \citep{mcintyre2022machine}.

One of the significant obstacles to using AI-based methods is the lack of credible data. \citep{su2021machine} generated a database on lift coefficients in microfluidic channels using a computationally expensive DNS method for three types of cross-sections. They then developed an MLP Neural Network to predict lift coefficients across 3 shapes of microfluidic channel cross-section. Using the same architecture, they trained the neural network on each cross-section shape separately. The provided model required to be retrained each time for inferring on new data.

In this work, we have trained a single network on the whole dataset provided by \citep{su2021machine}. We have studied the hyperparameters using a Bayesian optimization \citep{snoek2012practical} method with a hyperband \citep{li2017hyperband} stopping criteria to find the best architecture specific to the predefined need. We have also conducted a thorough study on the effects of Non-linearities and optimizers used. After validating the results with the experimental studies in the literature, we developed an API for the trained model with an easy-to-use user interface to reduce the need for machine learning expertise in order to use surrogate models for faster design iterations in inertial microfluidics applications. The developed webapp called microAI is deployed to the huggingface platform to be easily accessible to the community.

\section{Methods}
\subsection{Direct Numerical Simulation}
\label{sec:principle}
Point Particle Model (PPM) is widely adopted in the literature for determining particle trajectory \citep{bazaz2020computational} by removing the surface tension effect. Therefore, the lift force caused by the surface tension is not considered. The PPM assumption can be utilized when the particle size is considerably smaller than the channel dimensions. Otherwise, the effect of the lift force is not negligible because of the considerable pressure gradient and shear stress on the particle surface which makes the particles rotate in the flow. This rotation is one of the main reasons for the lateral displacement of the particles.

\begin{figure}[h!]
	\captionsetup[subfigure]{justification=Centering}
	\begin{subfigure}[t]{0.5\textwidth}
		\includegraphics[width=\textwidth]{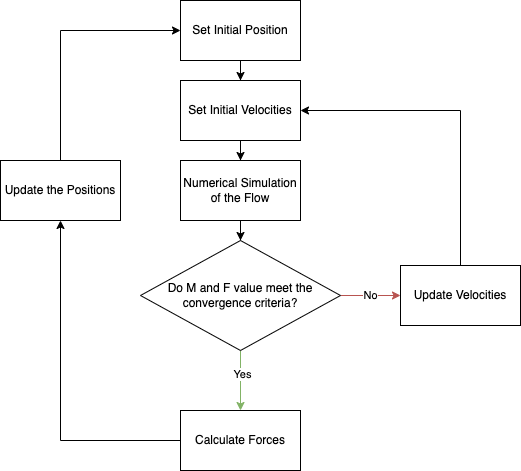}
		\caption{Flowchart showing the steps of the Direct Numerical Solution (DNS) method.}
	\end{subfigure}\hspace{\fill} % maximize horizontal separation
	\begin{subfigure}[t]{0.5\textwidth}
		\includegraphics[width=\linewidth]{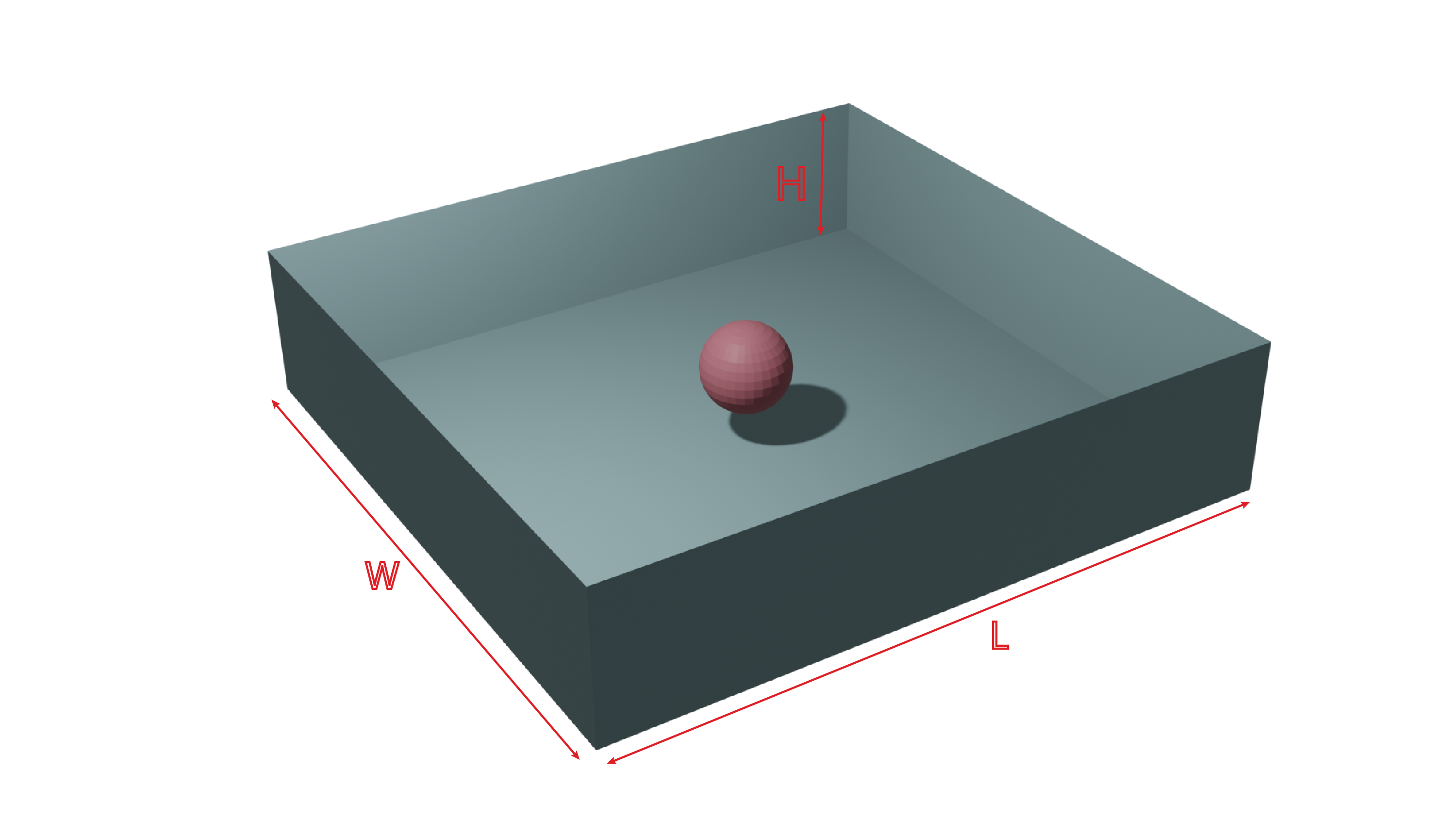}
		\caption{A particle with diameter d inside the microchannel.}
	\end{subfigure}
	\caption{Direct Numerical Solution method for calculating the lift coefficients inside a microchannel.}
	\label{fig:dns}
\end{figure}

In this study, a Direct Numerical Solution (DNS) is used to consider the effect of lift force even in PPM.
DNS method for lift force calculation has an iterative procedure that is shown in Figure \ref{fig:dns}. First, the initial coordinate of the particle is set and then, the initial velocities are set for the rotating sphere. Second, an FEM-based solution is used for determining the flow field and then the forces and moments applied to the particle are calculated. After calculating the linear and angular accelerations, the linear and angular velocity can be extracted by using an appropriate time step and they are used as an initial velocity for next iteration. This iterative process is continued while the calculated momentum in y and z direction and the force in the x direction decrease to less than $10^{-18}$ N.m and $1.5\times10^{-11}$ N, respectively. Finally, the lift force is calculated in the initial position that is a function of  the particle size ($d/H$) in which $d$ is the particle diameter and $H$ is the height of the channel, particle position inside the channel ($2y/H , 2z/H$) and Reynolds number of the channel by using the flow density $\rho$, maximum cross-section velocity $U_{max}$ and dynamic viscosity of the flow $\mu$, and channel height $H$ as the characteristic dimension of the channel ($\frac{\rho U_{max} H}{\mu}$). It should be noted that the particle center should always have a distance from the walls equal to its radius. Therefore, compared to PPM, the DNS method considers the volume of the particle, and the total lift force is calculable in this method.
A flowchart showing how the DNS method is used to calculate the lift force coefficients and the dimensions notation used in this work can be seen in Figure \ref{fig:dns}.

\subsection{Machine Learning}
\label{methods}

First, we describe the way that microAI could be used. After describing the dataset, we explain the network hyperparameters, the method used to sweep the hyperparameter space, and the final structure of microAI.

\subsection{microAI usage} 

The frequent need for lift coefficient for designing inertial microfluidic channels to calculate the particle trajectory resulted in the decision to have a fast, accurate, accessible, and easy-to-use application to save hundreds of hours from the scientific community.
The webapp was developed using gradio \citep{abid2019gradio}. Gradio is an open-source python library that facilitates the creation of API with methods and functions for adding input spaces such as text boxes, radio buttons, and sliders and removes the need for rewriting these functions from scratch. Another benefit of using gradio is its facility for deploying on the huggingface platform. Huggingface is a free and open-source hosting platform for machine learning models to bring them from just models in papers to functioning objects that can be accessed and used by the community. microAI was developed using pyTorch and can be accessed from this URL: \url{https://erfanhamdi.github.io/microAI/}

\subsection{Dataset}

\begin{figure}[h!]
	\centering
	\includegraphics[scale=0.5]{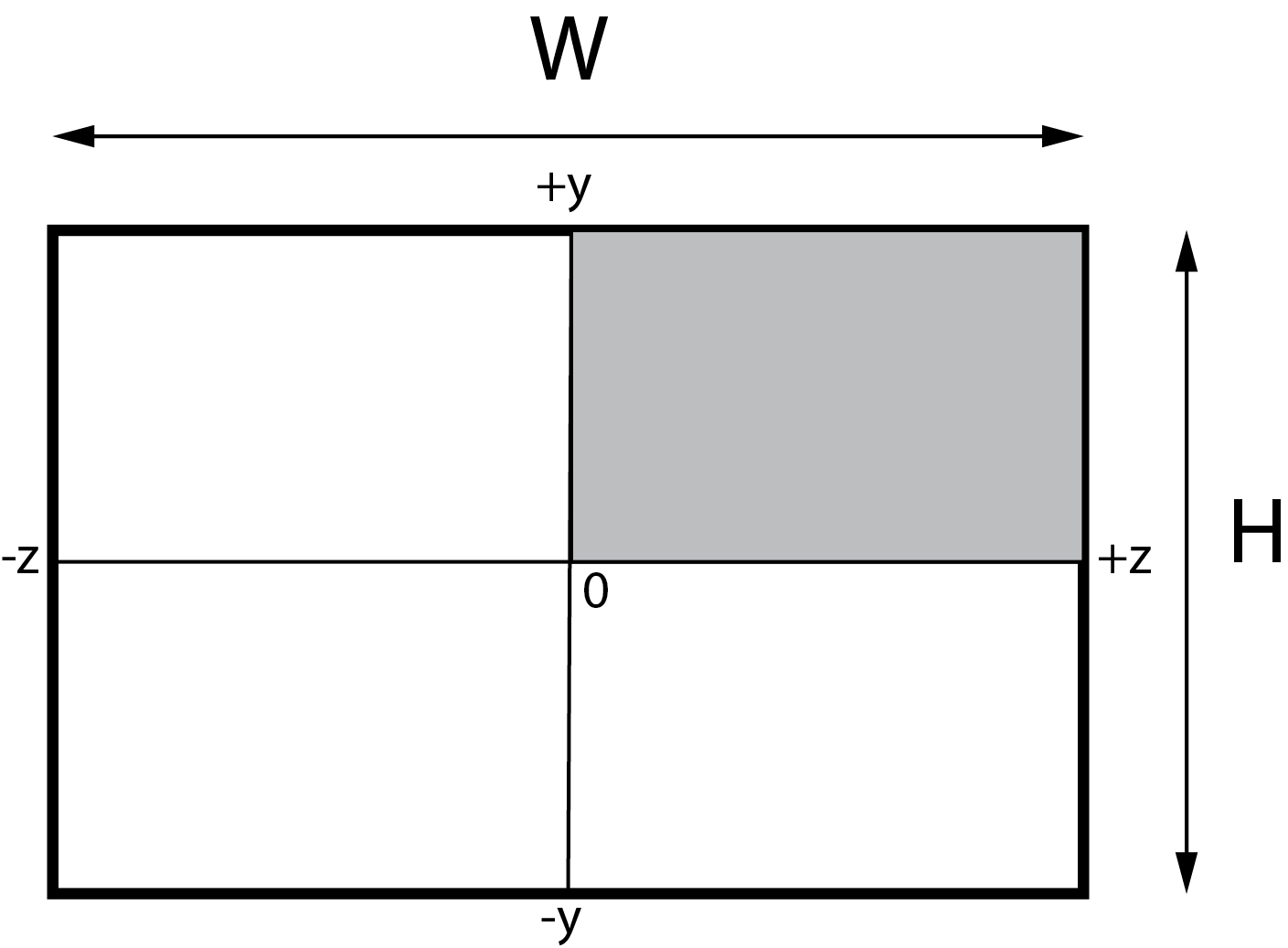}
	\caption{Rectangular cross section dimension annotations used for training the microAI}
	\label{fig:rect_guide}
\end{figure}

The dataset used for training the microAI was introduced by \citep{su2021machine}. This dataset comprises three parts, each for common shapes of microfluidic channel cross sections, Rectangular, Triangular and Semicircular. The input features of this dataset are channel aspect ratio (AR), Reynolds number of the flow (Re), blockage ratio ($\kappa$), and the collocation points across the cross section 2y/H and 2z/H. As the aspect ratio is only defined for rectangular cross sections, to be able to use only one network to infer the lift coefficients for each of these shapes, we set AR = 1 for the triangular and semicircular shapes. The shape of the cross section was also embedded inside the dataset using a one-hot encoding method resulting in a final eight-dimensional input dataset. The dimensions for the rectangular cross section and the first quarter that the lift coefficients are calculated for can be seen in the Figure \ref{fig:rect_guide}. 

This dataset, created by running a computationally expensive and accurate DNS method, offered lift coefficients for both y and z directions for selected points inside the cross-section domain.
In the original work, the training was done on each of the shapes separately. Thus the model could not use the data and information available in other shapes to have a better generalization on out-of-distribution data. 

\subsection{Model Optimization}

As machine learning models are highly dependent on their hyperparameters, which have a big space, it is necessary to have a search strategy to come up with the best set of hyperparameters that results in the desired performance.
There have been many efforts in devising a method for hyperparameter optimization, and one of the most used ones is the Bayesian hyperparameter optimization method. 
This model has many different hyperparameters. To optimize them, we have used a Bayesian optimization method utilizing the weights and biases platform \citep{wandb} to monitor and log the effect of each change with hyperband stopping criteria.
Bayesian methods model the conditional probability of the evaluation metric on a set of hyperparameters for a given black box model $p(y|\lambda)$. So that would result in a more computationally efficient search over the available space \citep{snoek2012practical}.
Hyperband hyperparameter optimization method \citep{li2017hyperband} is a specific form of random sampling method that, at the first step, trains the Network on a random sample of hyperparameter space for a few epochs and then keeps only the best performing ones for more prolonged training. The hyperparameter space was swept to reduce MSE on the test set. The range of the hyperparameters that were swept is in Table \ref{table:hp}. For this purpose data was splitted into train(80\%), validation(10\%) and test(10\%) sets. Training was done on the train set and the hyperparams were tuned using the performance of the model on the validation set. After the training, the final performance of the model was evaluated on the test set.

\begin{table}[h!]
	\caption{Hyperparameter range}
	\label{table:hp}
	\centering
	\begin{tabular}{ll}
		\toprule
		%		\multicolumn{2}{c}{Part}                   \\
		\cmidrule(r){1-2}
		Name      & Range \\
		\midrule
		Batch Size   & [256, 512, 1024, 2048]\\
		Learning Rate  & [1e-4, 0.01]  \\
		Hidden Layer Architecture & [256], [256, 128], [256, 512, ..., 128]\\
		\bottomrule
	\end{tabular}
\end{table}

At last, to select the best performing method, we used the model with the least MSE on the test set. As this webapp intends to reduce the computational cost of this calculation, the inference speed of the models was also accounted for. 

%\begin{figure}
%	\centering
%	\includegraphics[scale=0.05]{figs/bayesian_hyperband.png}
%	\caption{Hyperparameter Tuning using bayesian optimization method and Hyperband stopping criteria}
%\end{figure}
\subsubsection{Neural Network Optimizer}
\begin{figure}[h!]
	\centering
	\includegraphics[scale=0.5]{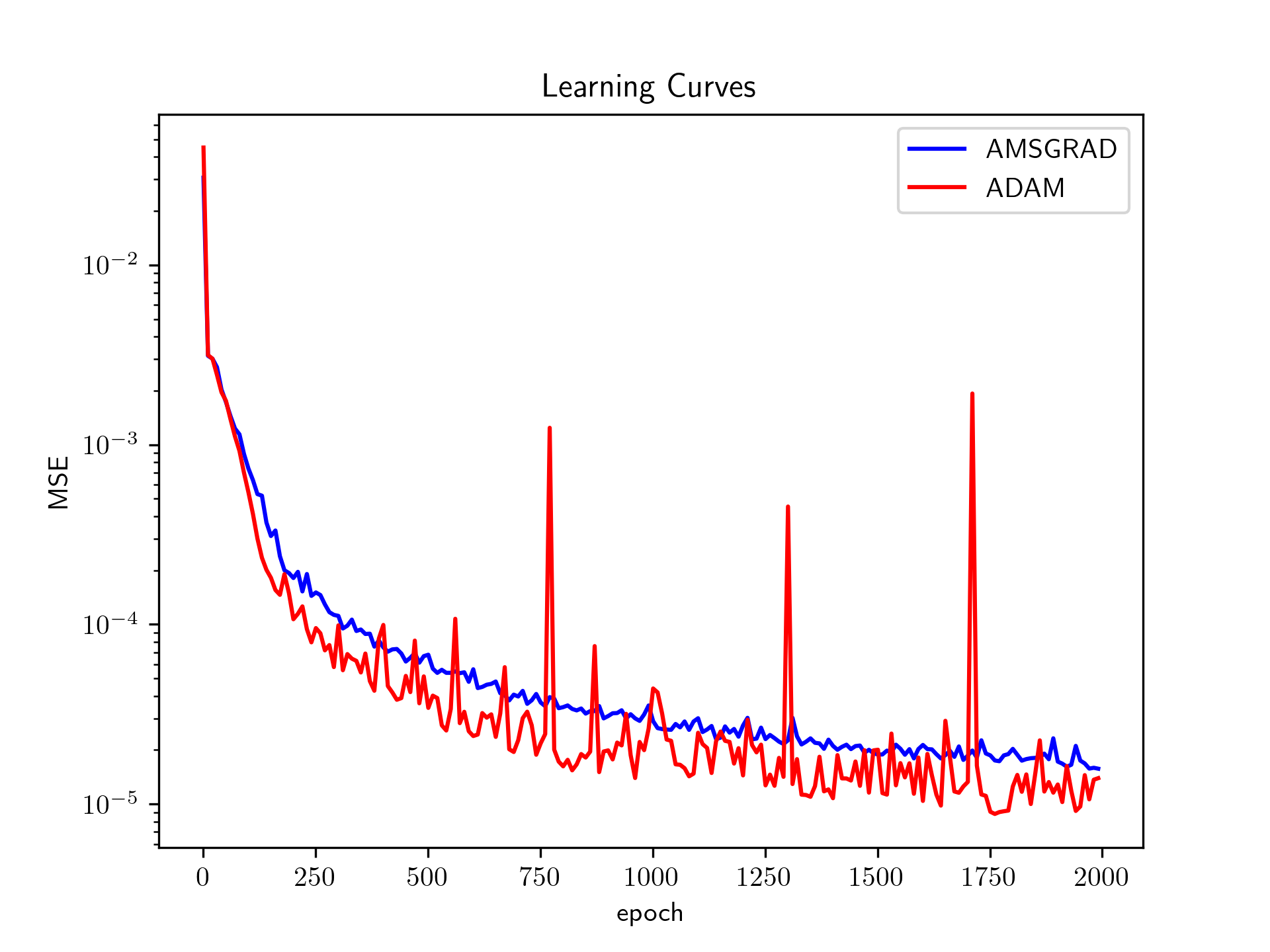}
	\caption{Training convergence comparison between ADAM and ADAM with AMSGRAD}
		\label{fig:amsgradVadam}
\end{figure}

Many optimization methods have been offered in the literature for training neural networks using stochastic gradient descent (SGD) as their core. Other methods, such as ADAGRAD \citep{duchi2011adaptive} and RMSPROP \citep{hinton2012neural}, have added a scaling property to the SGD that helps update the learning rate automatically based on the history of the gradients. This scaling method has resulted in much better performance than the SGD in different applications. The first optimization algorithm that offered this scaling method was ADAGRAD. Other methods built upon ADAGRAD and solved various problems, such as rapid learning rate decay in high dimensional problems. ADAM \citep{kingma2014adam} mitigated that problem by using the exponential moving average of the squared past gradients. This resulted in significant improvements in the convergence of several practical applications, but they sometimes failed to converge. \citep{reddi2019convergence} diagnosed that the problem was caused by the relatively quick decay of large gradients because of exponential averaging and proposed a variant of ADAM that uses a long-term memory of the past gradients with the same time and space requirement as the original ADAM called AMSGRAD. To achieve the required MSE error on this dataset, we first tried using ADAM optimizer but struggled with converging and stability. AMSGRAD resulted in a more stable convergence while performing the hyperparameter optimization step. The results of changing the optimizer from ADAM to AMSGRAD can be seen in Figure \ref{fig:amsgradVadam}. As can be seen, AMSGRAD has more stable performance training the network.

\subsubsection{Activation Functions}
\begin{figure}[h!]
	\centering
	\includegraphics[scale=0.4]{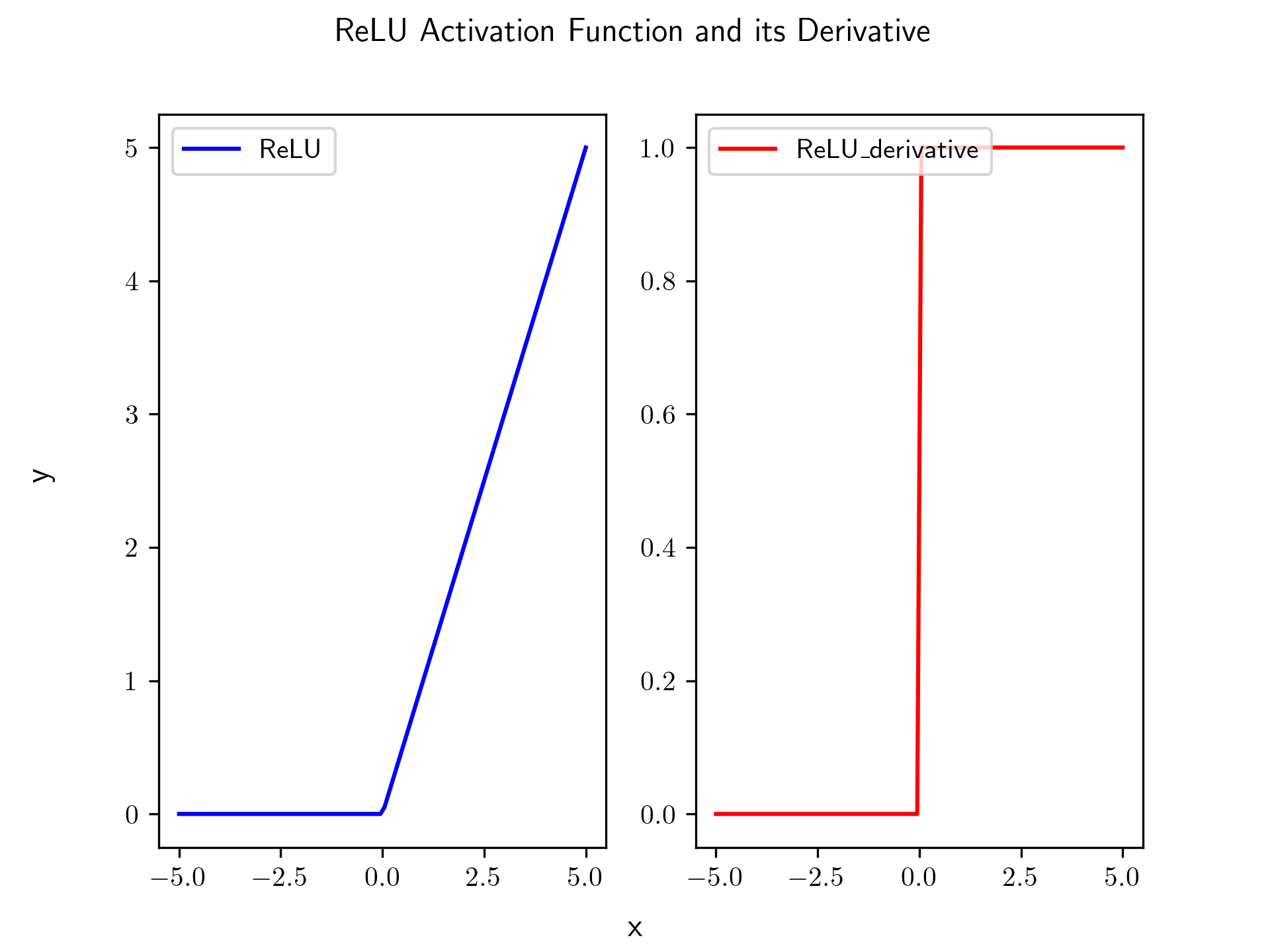}
	\caption{ReLU activation function and its' derivative}
	\label{fig:relu}
\end{figure}

\begin{figure}[h!]
 	\centering
 	\includegraphics[scale=0.4]{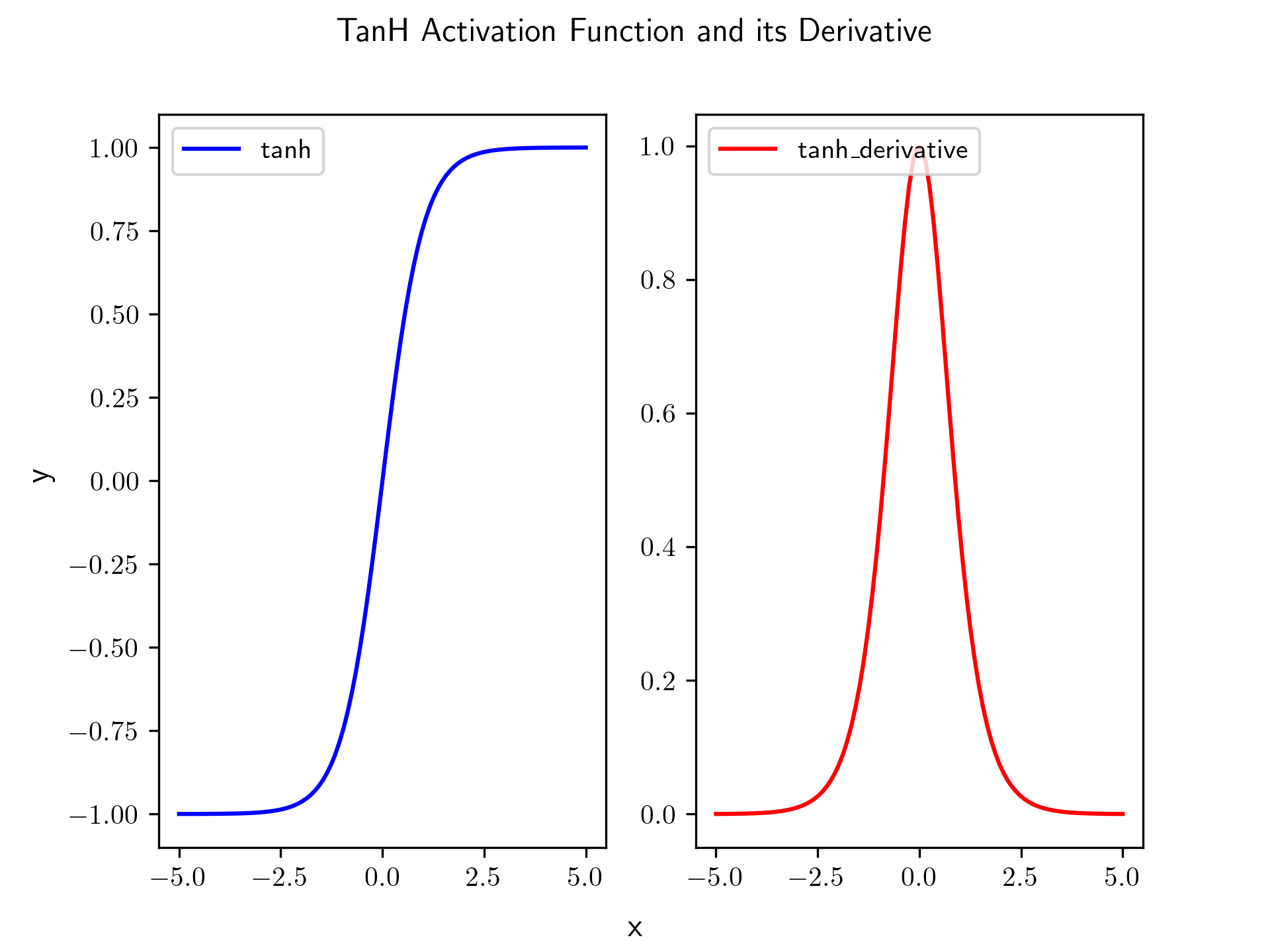}
 	\caption{ReLU activation function and its' derivative}
 	\label{fig:tanh}
 \end{figure}

Activation functions add non-linearity to the neural networks, and there have been many suggestions for activation functions. Two of the mainly used activation functions in the literature are ReLU and Hyperbolic Tangent.
 Rectified Linear Units \citep{glorot2011deep} are the most widely adopted activation functions in machine learning, especially in problems having sparse data. Neural Networks that use ReLU as the activation function are known to optimize faster because optimizing linear functions can be easier \citep{goodfellow2016deep} But  as can be seen by the Equation \ref{eq:relu}, ReLU shows a linear response in the positive input range and it is not differentiable in zero which can be seen in Figure \ref{fig:relu}.
\begin{equation}
	f(x) = max(0, x)
	\label{eq:relu}
\end{equation}
As in our application, the lift coefficients may have positive or negative values, using ReLU would not help to learn the underlying function faster in smaller networks. Another widely used activation function is hyperbolic tangent (TanH) that can be seen in Equation \ref{eq:tanh}. It is computationally more expensive than ReLU but is differentiable and symmetric around zero and saturated. TanH function and its derivative can be seen in Figure \ref{fig:tanh}. As can be seen, ReLU has a zero gradient in the regions below zero, which means that it would result in the network not being able to learn on the negative values, but TanH is symmetric around zero. Which would help the model to learn better on this specific application.
 \begin{equation}
 	f(x) = \frac{2}{1+e^{-2x}} - 1
 	\label{eq:tanh}
 \end{equation}
 These properties make it a better choice for regression applications such as this one where the answer should be smooth and differentiable everywhere.
 
\begin{figure}[h!]
	\centering
	\includegraphics[scale=0.15]{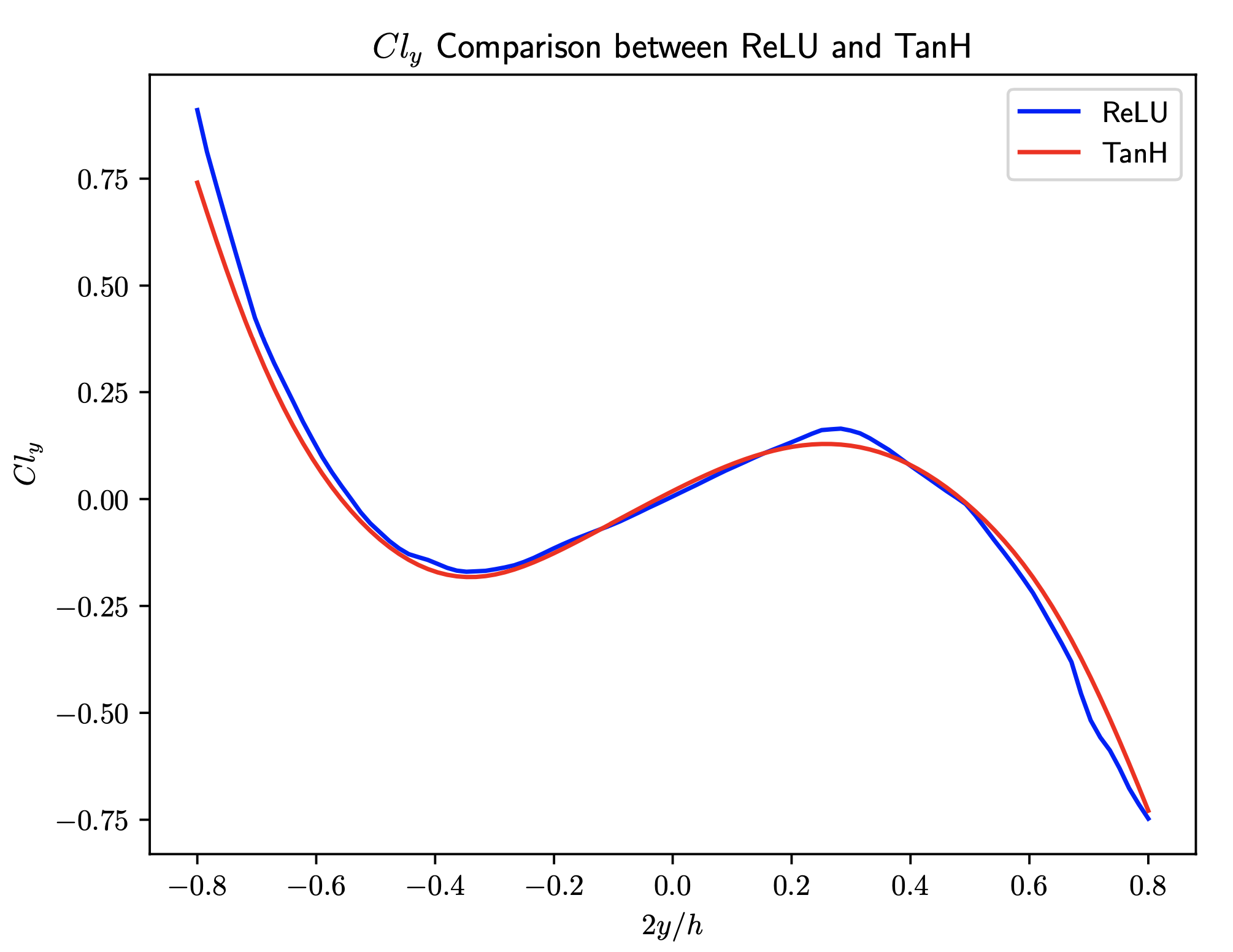}
	\caption{Comparison between the final model inference using Tanh and RelU activation function.}
		\label{fig:reluVtanh}
\end{figure}

 We then trained our network using both of these activation functions with fixing every other parameters and let them both reach to the predefined MSE metric and then compared the results on a test case for a semi-circular cross-section. The results can be seen in Figure \ref{fig:reluVtanh}. There were no obvious difference in the convergence rate between these two functions but the model trained using TanH as the activation function resulted in a smoother function.
\subsection{Final Model Training}
We selected the final model architecture by the result of the hyperparameter sweep and trained the final model for 2000 epochs to reach the $1.5\times10^{-5}$ MSE error on the test set which was reported by \citep{su2021machine}. The set of final hyperparameters selected can be seen in Table \ref{table:final_hp}.
\begin{table}[h!]
	\caption{Final hyperparameters}
	\label{table:final_hp}
	\centering
	\begin{tabular}{ll}
		\toprule
		%		\multicolumn{2}{c}{Part}                   \\
		\cmidrule(r){1-2}
		Name      & Range \\
		\midrule
		Epoch   & 2000  \\
		Batch Size   & 256\\
		Learning Rate  & 0.004172 \\
		Hidden Layer Architecture & [256, 512, 128]\\
		Optimizer & AMSGRAD\\
		Activation Function & TanH\\
		Loss Function & MSELoss\\
		\bottomrule
	\end{tabular}
\end{table}
The input data was normalized in (0, 1) range using MinMax scaling. There was no apparent difference between normalizing using different methods, such as Standard Scaling. 
\section{Results}
Simulating particle trajectory before fabrication is necessary to avoid wasting time and materials. At first the microAI results were verified by comparing the produced lift coefficients from the DNS results and then the result of two particle trajectory simulations are compared with experimental results. The first experiment is on, a multi-cross-section channel including rectangular, triangular, and semi-circular shapes is investigated for the purpose of single particle focusing. Then particle trajectory inside a serpentine channel containing more complicated flow patterns like secondary flows is studied.
To compare the performance of microAI with the DNS code, we have compared the results with the data available in the literature \citep{kim2016inertial, lu2015inertia, wang2017analysis} and the results of DNS calculations provided by \citep{su2021machine}. 

\begin{figure}[h!]
	\captionsetup[subfigure]{justification=Centering}
	\begin{subfigure}[t]{0.45\textwidth}
		\includegraphics[width=\textwidth]{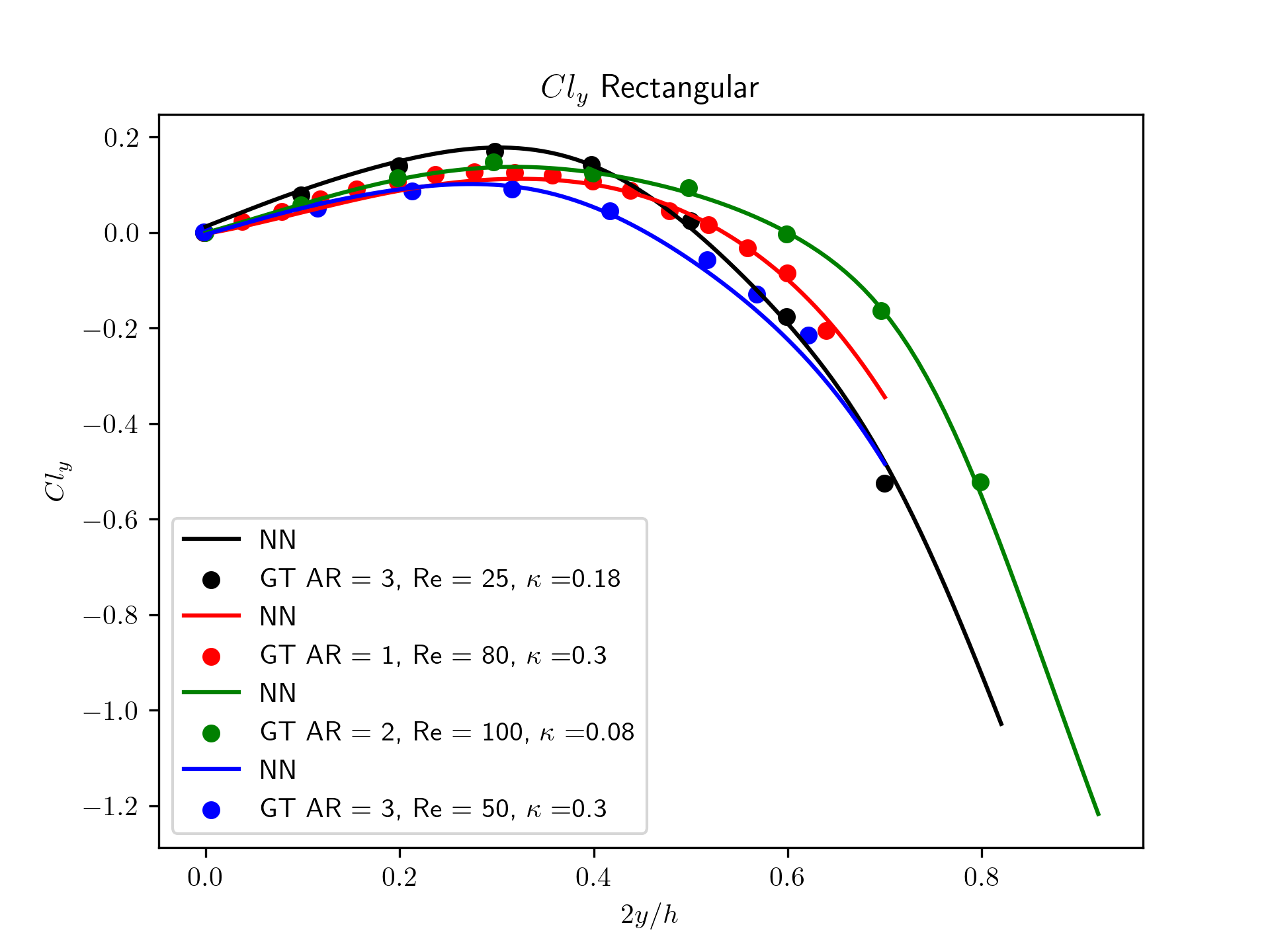}
		\caption{$Cl_y$ for the centerline of the rectangular cross section. Ground Truth (GT) data reference for the (AR=3, Re=25, $\kappa$=0.18)\citep{su2021machine}, (AR=1, Re=80, $\kappa$=0.3)\citep{lu2015inertia}, (AR=2, Re=100, $\kappa$=0.08)\citep{su2021machine}, (AR=3, Re=50, $\kappa$=0.3)\citep{kim2016inertial}}
	\end{subfigure}\hspace{\fill} % maximize horizontal separation
	\begin{subfigure}[t]{0.45\textwidth}
		\includegraphics[width=\linewidth]{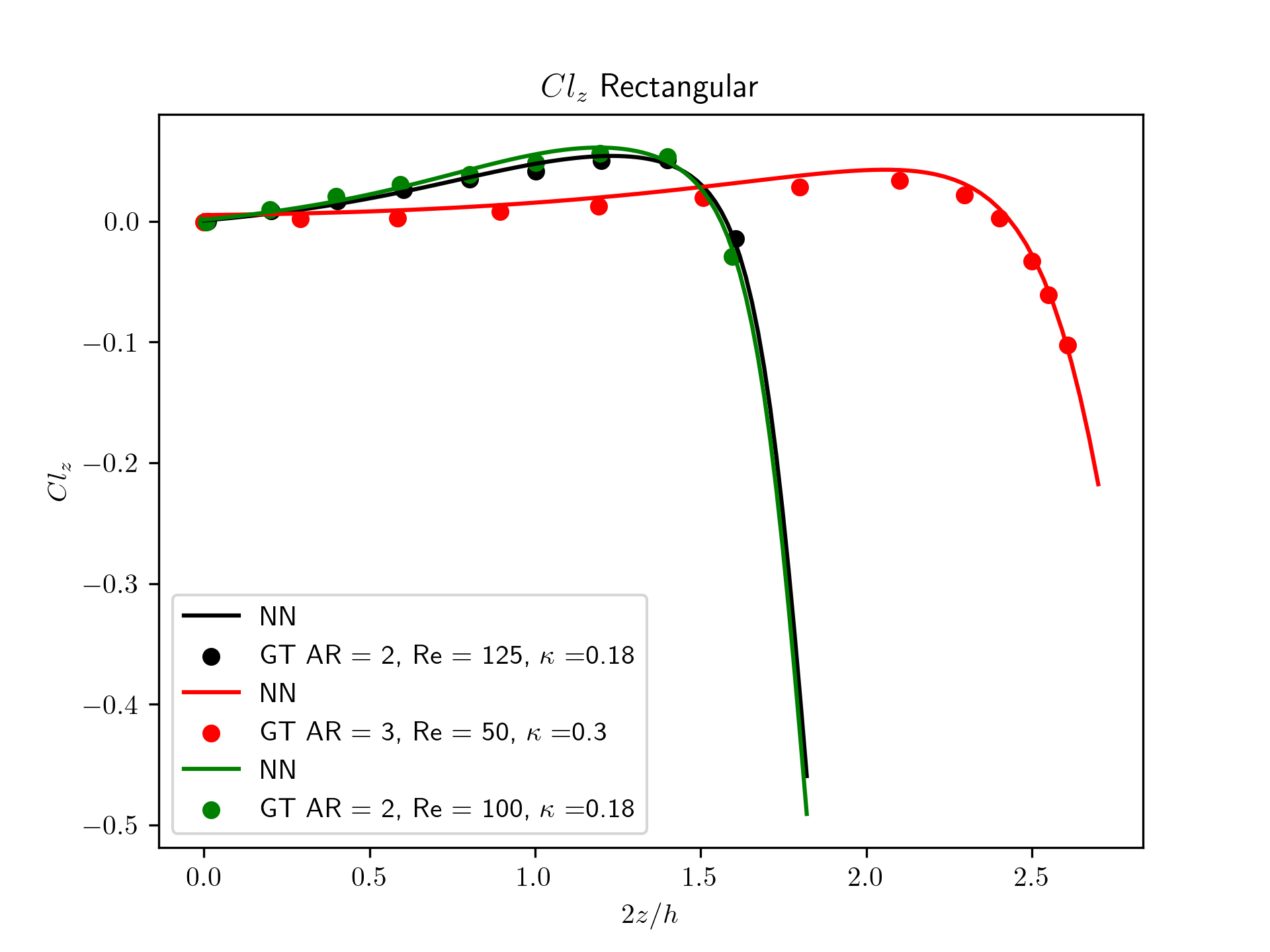}
		\caption{$Cl_z$ for the centerline of the rectangular cross section. Ground Truth (GT) data reference for the (AR=2, Re=125, $\kappa$=0.18)\citep{su2021machine}, (AR=3, Re=50, $\kappa$=0.3)\citep{kim2016inertial}, (AR=2, Re=100, $\kappa$=0.18)\citep{su2021machine}}
	\end{subfigure}
%	\bigskip % more vertical separation
	\begin{subfigure}[t]{0.45\textwidth}
		\includegraphics[width=\linewidth]{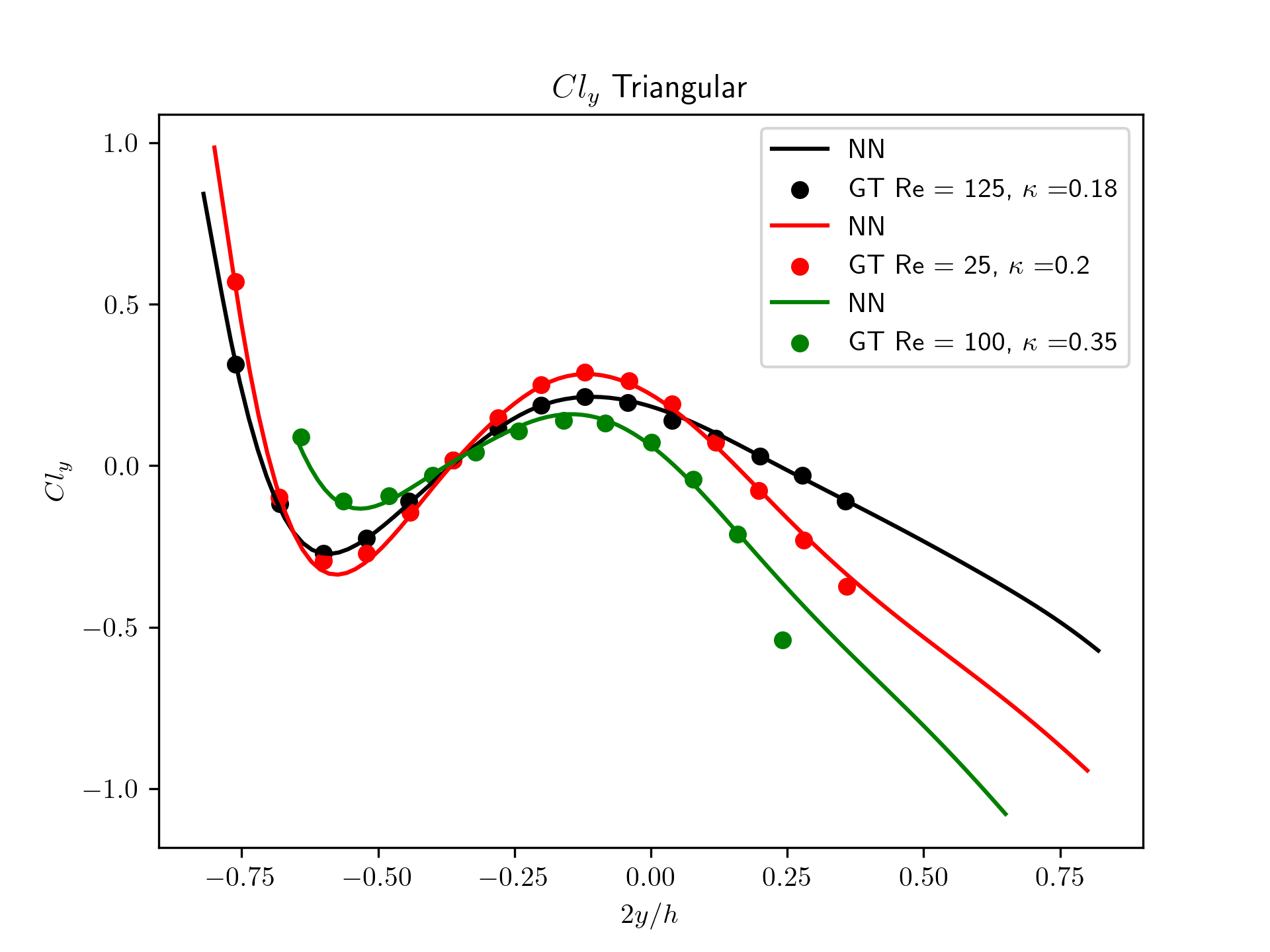}
		\caption{$Cl_y$ for the centerline of the triangular cross section. Ground Truth (GT) data reference for the triangular cross section is from DNS calculations by \citep{su2021machine}}
	\end{subfigure}\hspace{\fill} % maximize horizontal separation
	\begin{subfigure}[t]{0.45\textwidth}
		\includegraphics[width=\linewidth]{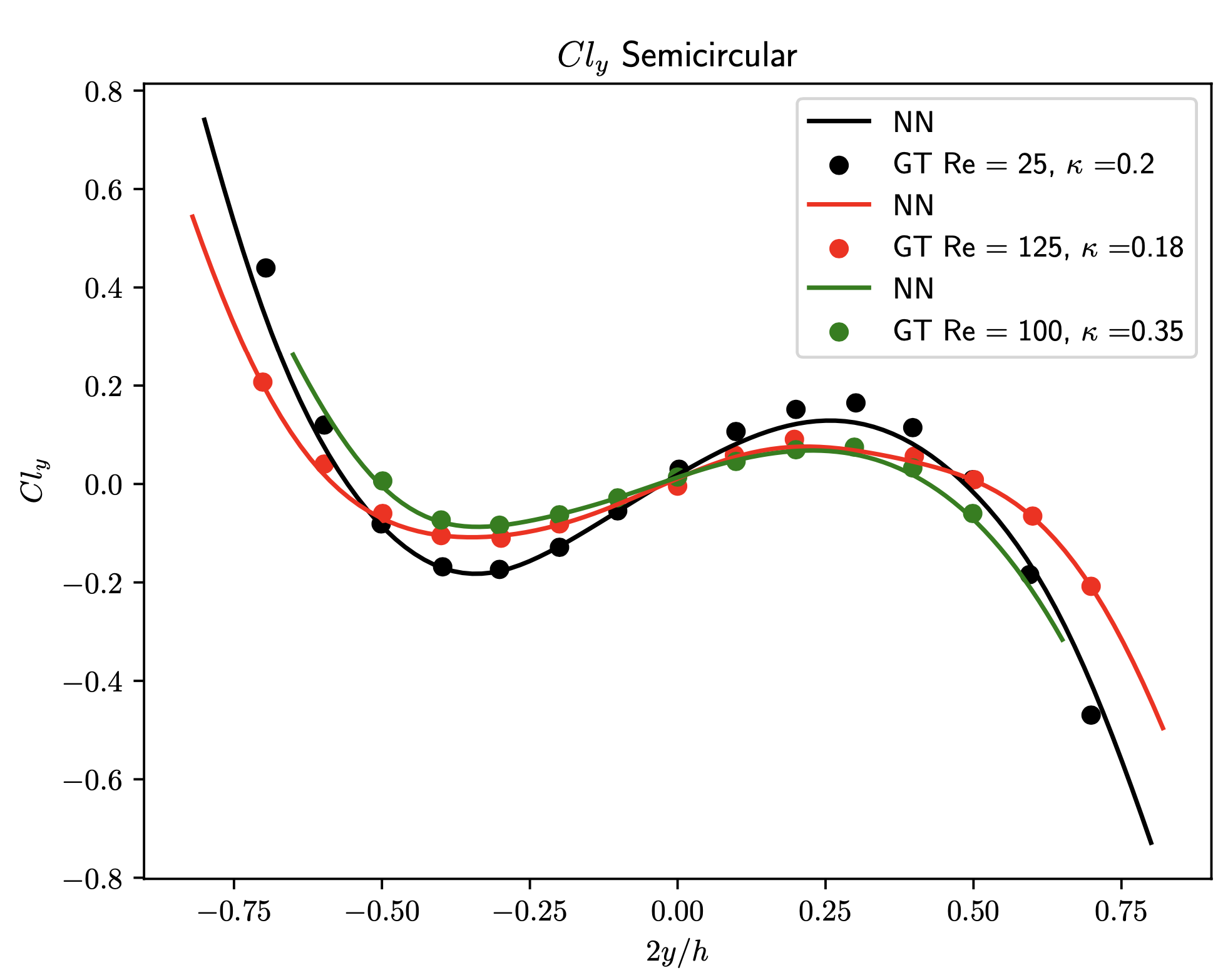}
		\caption{$Cl_y$ for the centerline of the semicircular cross section. Ground Truth (GT) data reference for the triangular cross section is from DNS calculations by \citep{su2021machine}}
	\end{subfigure}
	\caption{Comparison of the microAI with the lift coefficients reported in the literature.}
\label{fig:res}
\end{figure}

The resulting Lift coefficients for each cross-section and their comparison with the ground truth is in Figure \ref{fig:res}. As can be seen, microAI shows good compliance with the ground truth. Figures \ref{fig:res}-a, \ref{fig:res}-c and \ref{fig:res}-d show the y component of the lift coefficient along the centerline in $y$ direction and Figure \ref{fig:res}-b shows the $z$ component of the lift coefficient of a rectangular cross section along the centerline in $z$ direction. Each Figure contains an extrapolation case as well as interpolation. An important performance metric of interest in the application of machine learning methods to solve mechanical engineering problems is the out of distribution performance of the machine learning model \citep{yuan2022towards}. As can be seen in the Figure \ref{fig:res} microAI shows a good performance on test data out of the training distribution.

\begin{figure}[h!]
	\centering
	\includegraphics[height=200pt]{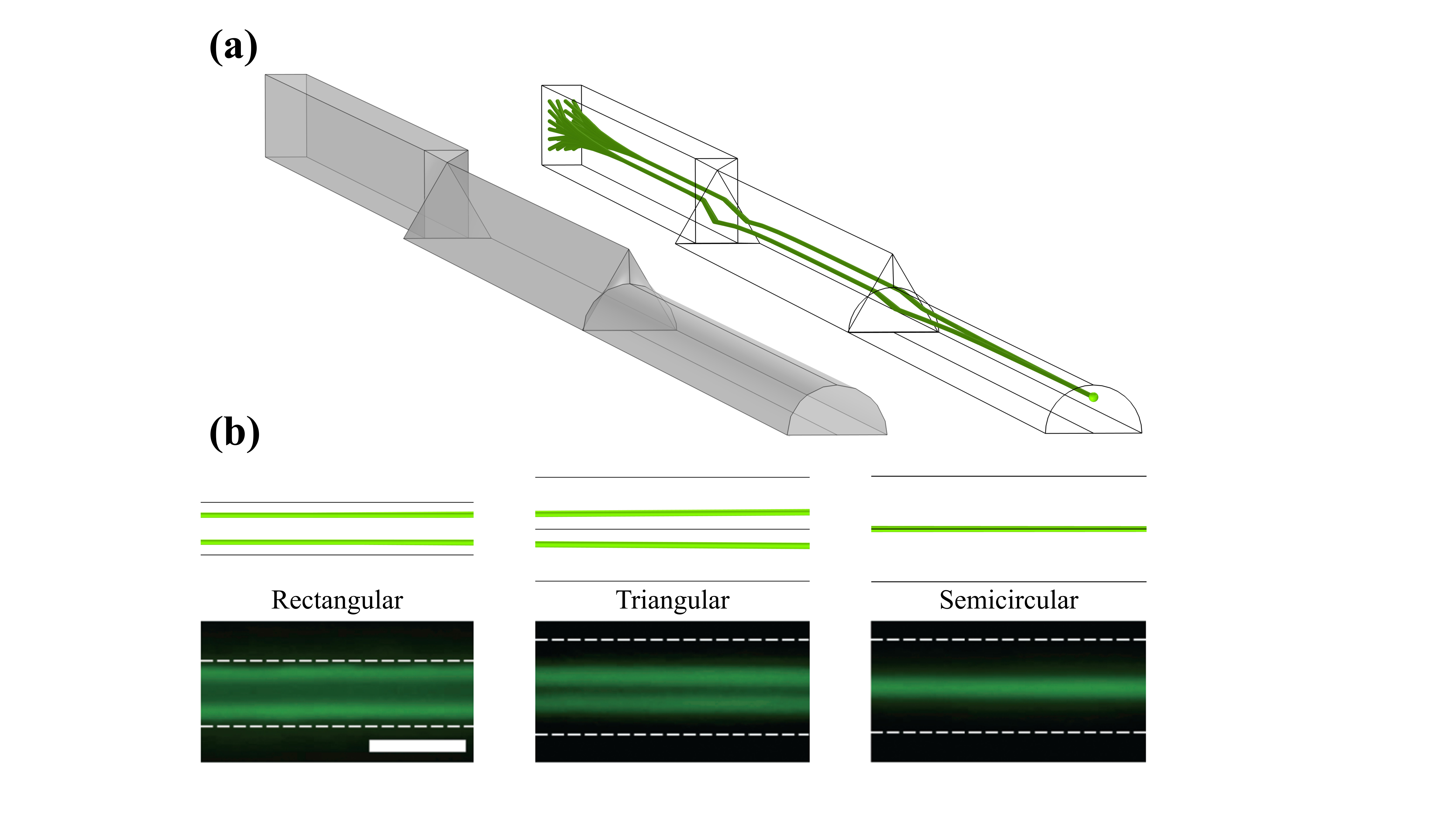}
	\caption{The validation study schematic includes all three cross-sections investigated in the current study. a) Particles pass through a channel with various channel cross-sections, including rectangular, triangular, and semi-circular, respectively. Particles are focused in a line due to the lift forces applied by the channel's geometry. b) Illustration of particle trajectory in the rectangular, triangular and semi-circular cross-sections to compare numerical results and experimental study by Kim et al. (reproduced from \citep{kim2016inertial} with permission from the Royal Society of Chemistry). The white scale bar represents 50 $\mu m$.
	}
	\label{fig:3particle}
\end{figure}

As microAI is intended to facilitate the simulations regarding the design of Inertial microfluidic systems, it is important to be able to validate its performance with application on reproducing benchmark simulations and experiments.
\FloatBarrier
Two benchmark experimental case studies have been inspected to validate the results further. In the first one, a straight channel comprised of three different shapes of cross-sections have been simulated using the lift coefficients calculated by microAI. This channel consists of a 2.5 cm long rectangular cross section, with AR = 0.5, and Re = 135.5 and $\kappa$ = 0.1. In order to calculate the lift coefficients for this cross section, we calculated the lift coefficients for a rectangle with AR=2 and switched the columns for $CL_y$ and $CL_z$. The next section is 1 cm long channel with a triangular cross section with the Re = 175 and $\kappa$ = 0.1. And the last part is a 1 cm long channel with semi-circular cross section with the Re = 89.635 and $\kappa$ = 0.2. Using DNS calculation would not be feasible for doing any iterations in the design of the microchannel but we can calculate the Lift coefficients using microAI and map it to the channel to use it further in particle tracking simulations. In the first section, as shown in Figure \ref{fig:3particle}-a, the particles motion starts from the rectangular cross-section evenly distributed and passes a 2.5 cm long channel; particles focus near the side walls during this path. Then, particles maintain the same equilibrium positions at a lower height when moving in the triangular channel. Finally, two equilibrium positions change to a single position pattern inside the semi-circular channel. The results are compared to the experiment conducted by \citep{kim2016inertial} as can be seen in Figure \ref{fig:3particle}-b, the separation of particles is in good compliance with the experiment with varying cross-sectional shapes.

\begin{figure}[h!]
	\centering
	\includegraphics[height=180pt]{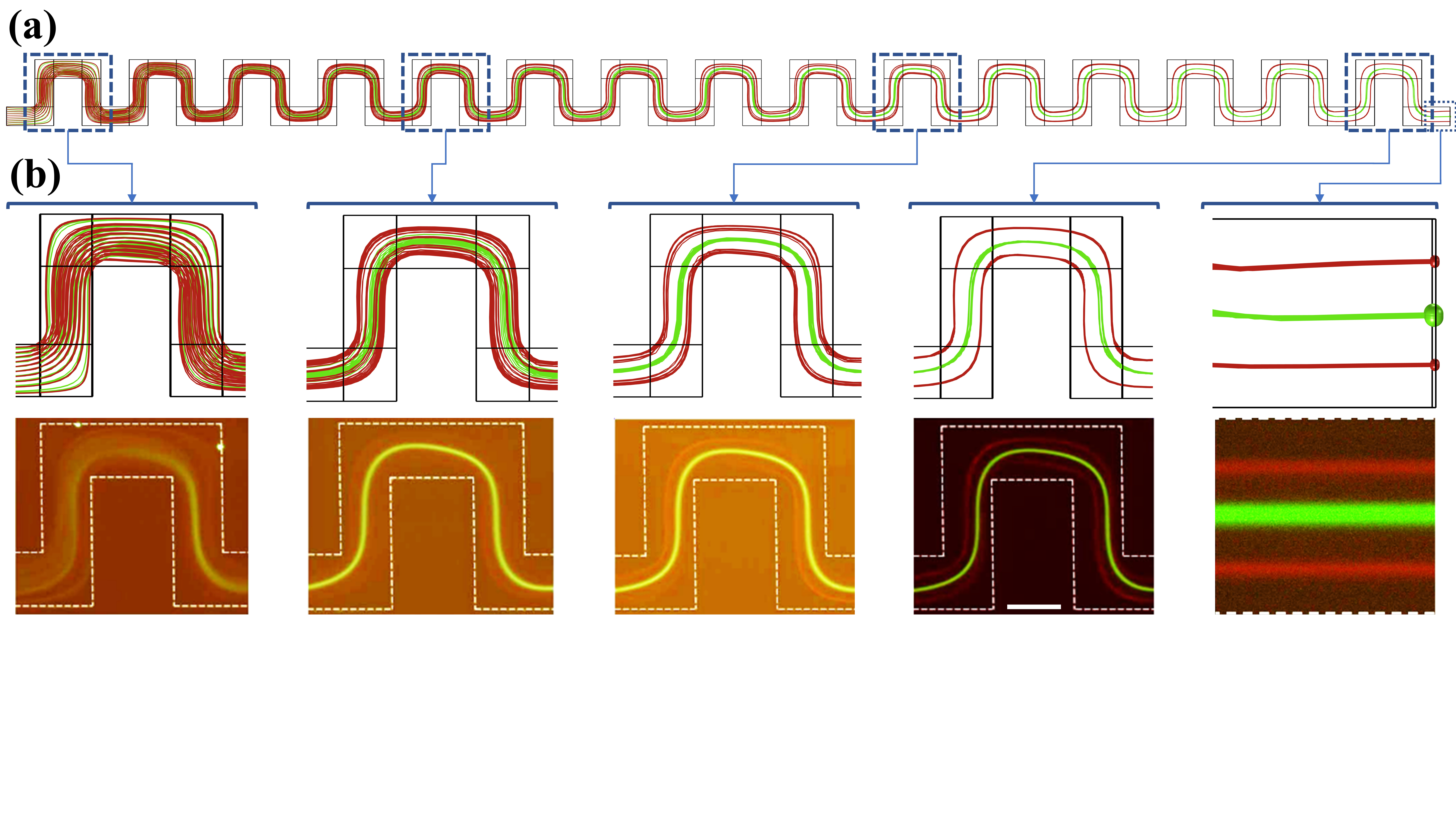}
	\caption{Particle separation using a serpentine channel with a rectangular cross-section. a) Particles entered from all regions of a rectangular cross-section and focused in their equilibrium positions thanks to the lift force. b) the numerical results are compared with \citep{zhang2014inertial}’s experimental study in different parts of the channel, and the separation results are shown at the end (Reproduced from \citep{zhang2014inertial} with permission from Nature Publishing Group). The white scale bar represents 200 $\mu m$.}
	\label{fig:serpentine}
\end{figure}

Particle separation is another application of inertial microfluidics. Figure \ref{fig:serpentine}-a indicates a serpentine channel in which 10 $\mu m$ and 20 $\mu m$ particles show different behaviors because of the difference in their lift force. The only unequal parameter is $\kappa$ which is 0.25 and 0.5 for 10 $\mu m$ and 20 $\mu m$ particles, respectively. Like the other validation, 2y/H and 2z/H are selected across the cross-section with the constant step of 0.02. In addition, Re and AR were chosen 34.56 and 5 as the input to microAI. Figure \ref{fig:serpentine}-b compares the experimental study by \citep{zhang2014inertial} and the simulation using lift coefficients by microAI. The red lines indicate 10 $\mu m$ particles, and the green lines represent 20$\mu m$ particles. Five sections of the channel are shown that verify the validity of the numerical results. As shown in Figure \ref{fig:serpentine}-b, particles enter the channel without any specific order. After passing through the serpentine channel, they find their equilibrium position, and finally, particles will be separated. The final equilibrium position in the serpentine channel is not only dependent on the lift force (similar to previous validation), but also the drag force resulting from the secondary flow has an essential effect on particles. Therefore, the final position is where lift and drag forces are in equilibrium.
%\FloatBarrier
%\pagebreak
\subsection{Conclusion}
In this work we have introduced microAI which is a webapp used for fast calculation of the lift coefficients in microchannels with three shapes of cross section to make the design iterations for microfluidics systems faster, easier and less reliant to expert knowledge. We have also studied the impact of choosing optimizer and activation function on training convergence and the final function smoothness which is required for this kind of applications. It seems that using ReLU for applications involving mechanical engineering problems would not result in good performance. In order to validate the performance of microAI we have extracted the lift coefficients using microAI and simulated two common applications of inertial microfluidics, including particle focusing and particle separation. Particle trajectories were validated successfully with previous experimental studies.

%\bigskip
%\newpage
\bibliographystyle{unsrtnat}

\end{document}